\documentclass[12pt, preprint]{aastex}
\usepackage{graphicx}

\received{} \accepted{}

\shortauthors{Y. P. Wang et al.} \shorttitle{Star formation
history associated with QSO activities}

\slugcomment{to submit to Astrophysical Journal}

\begin{document}

\title{Cosmic star formation history associated with QSO activity: an approach using the black hole to bulge mass correlation}

\author{Y. P. Wang$^{1,2}$, T. Yamada$^2$, Y. Taniguchi$^3$}
\affil{$^1$Purple Mountain Observatory, Academia Sinica, China;
National Astronomical Observatories of China} \affil{$^2$National
Astronomical Observatory of Japan, Mitaka, Japan}
\affil{$^3$Tohoku University, Aramaki, Aoba, Sendai 980-8578,
Japan}

\begin{abstract}
The tight correlation between the masses of central black
holes and their host spheroids in nearby galaxies and active
galactic nuclei (AGN) suggests that black hole growth is
closely related to their spheroid formation. Based on our previous
work regarding such a joint evolutionary scheme and the consequential
black hole to bulge mass correlation, we use the X-ray luminosity function of AGN and
the cosmological evolution rate which are from ROSAT X-ray surveys to estimate
the cosmic star formation history associated with the
black hole growth. By the basic assumption that the major black hole growth occurs
during the luminous AGN phase, the luminosity
function of AGNs as a function of redshift traces not only the
accretion history of the black holes but also the cosmic star
formation history of the spheroids.

Although the space density of the especially luminous QSOs is very
low, we show that the total amount of star formation associated with the massive black
hole growth is almost the same as that of Lyman Break Galaxies
detected by the current optical deep surveys. We thus argue
that the optical deep surveys may miss about half of the net
star formation in our Universe. This is probably due to in part significant
dust extinction as well as
the small field of view of previous optical surveys which cannot
sample such rare events with relatively short time scale.
However, the far infrared emission from the dust heated by star
formation on-going during the black hole growth could sufficiently
account for the observed SCUBA number counts, and would be the probable 
dominating energy source of the SCUBA population.

\end{abstract}


\section{Introduction}
Recent observations reveal a tight correlation between the
masses of central black holes and their host spheroids in nearby
galaxies and AGNs \citep{Kor95, Fab97, Mag98, Lao98, Ho99, Wan99a}. 
A mean value of the black hole to bulge mass ratio, $\rm M_{\rm bh}/M_{\rm
bulge}\,\sim\, 10^{-3}$ and the local black hole mass density,
$\rho_{\rm bh}\,\sim\, 3 \,-\, 5\,\times\, 10^5\,\,{\rm M_{\odot}}/{\rm Mpc^3}$ are now fairly
agreed upon by observations using different techniques \citep{Sal99,
Kor01, Mer01, McL02}. This correlation implies a possible scenario
where the black hole accretion history in the inner few parsecs
and the $\,\sim\,\rm kpc$ scale star formation during spheroid formation
are closely connected. Simply speaking, the amount of gas
accreted to grow the central black hole is proportional to
the amount of gas driving the star formation to populate the
spheroid within a similar time scale of the two activities \citep{Wan98, Wan00, Hae00, Mon00, Bur01, Pag01}.

This relation also opens an important avenue to the understanding of
the cosmic star formation history (CSFH) associated with
the black hole growth by accretion, complementary to the previous
optical/near-infrared (NIR) or far-infrared deep surveys.

The study of the CSFH from the optical/NIR deep surveys describes
only the unobscured star formation in galaxies, and may
be blind to activities during the massive
spheroid formation because of the significant dust extinction or the small field of view \citep{Mad96, Ste96, Con97,Ste99}. In fact, our knowledge
of the cosmic star formation history, especially at high redshift, has
dramatically changed in the last few years due to the results
of the far-infrared and submillimeter deep surveys. The compelling lines of
evidence show that the absorption and re-radiation of light by
dust in the early epoch of galaxy formation and evolution is
significant \citep{Pug96, Gui97, Sch98, Fix98, Hau98, San99,
Bar99}.

Although the observations at far-infrared and submillimeter wavelengths may
help to unveil the early dusty star formation epoch, it is
still a very difficult job to obtain an unbiased view of the early star
formation history from these luminous infrared systems due to the rather poor
angular resolution of SCUBA and to the faintness of the optical
counterparts \citep{Bla99a,DeZ01}.

The main purpose of this paper is to fairly reconstruct the global
star formation history associated with AGN accretion and
give a proper estimation of the star formation activity during 
spheroid formation. We assume the X-ray luminosity $\rm L_{\rm x}$
of the AGN is powered by accretion onto a central massive black hole,
especially during the luminous phase. Deep X-ray surveys provide a direct
probe of the AGN accretion history, and the history of the black hole
growth as well as the joint star formation, based on the
assumption that the luminous AGNs reflect the stage of major black
hole growth and spheroid formation \citep{Boy00, Miy00}.  In this
paper, we trace the AGN evolution with the soft X-ray $0.5\,-\, 2\,\,{\rm keV}$ local
luminosity function and the evolution rate by Miyaji et al.
(2000), which gives an excellent fit to the ROSAT surveys of various depth,
including the number counts and the soft X-ray background.

The X-ray view of the CSFH is recently
discussed by several authors for normal spiral galaxies where the
X-ray emission is dominated by a population of X-ray binaries, hot
interstellar gas, or even low luminosity AGNs (LLAGNs) with fluxes about
$10^{-17}\,-\,10^{-18}\,\,{\rm erg\,\,cm^{-2}\,\,s^{-1}}$ at $0.5\,-\,2\,\,{\rm keV}$ 
energy band \citep{Cav00, Pta01, Hor02, Miy02}. It is not the purpose of this
paper to give any constraints on the star formation in galaxies
or LLAGNs. Instead, we study
the intensive star formation associated with black hole growth in massive spheroids
using the X-ray surveys. We adopt in the calculation: 1) a
simple connection between the X-ray emission from AGNs and the black
hole mass by assuming an Eddington ratio $\epsilon$; 2) the black
hole to bulge mass correlation; 3) a similar time scale for the
black hole growth and the intensive star formation populating
finally the spheroids based on our previous work. The present day black hole density
and the SCUBA number counts are used as two important model
constraints in the calculation, with the set of cosmological
parameters $(\Omega_{\rm m}, \Omega_{\rm \Lambda})=(0.3, 0.7)$ and
$\rm H_{0}\,=\,50\,\,{\rm km/s/Mpc}$.

\section{Black hole mass distribution and the accretion history from ROSAT surveys\label{sec2}}

The mass accretion onto the central black hole is an efficient source of
X-ray radiation, especially for the luminous AGNs.
Although X-ray binaries can contribute as much as $\,\sim\,
5\%\,-\,10\%$ of the X-ray background (XRB) flux in the $0.5\,-\,2\,\,{\rm keV}$ band, the bulk of
the energy density of the XRB is certainly explained by AGNs \citep{Mus00, Bar01a, Gia01}. Thus, the deep X-ray surveys could
present good opportunities for the study of accretion history of
AGNs. We estimate the black hole mass from the X-ray luminosity by
assuming an Eddington ratio $\epsilon$, which is defined as 
the fraction of Eddington luminosity at which AGN radiate 
($\epsilon\,=\,\frac{\eta\,\dot{m}\,c^2}{\rm L_{\rm Edd}}$). With the bolometric
correction $\beta$ ($\rm L_{\rm bol}\,=\,\beta\,L_{\rm x}$), we find
the black hole mass $\rm M_{\rm bh}=\frac{\beta\,L_{\rm
x}}{0.013\,\epsilon}$. $\rm L_{\rm x}$ is the AGN $0.5 \,-\, 2\,\,{\rm keV}$
luminosity in units of $10^{40}\,\,{\rm erg\,\,s^{-1}}$ and $M_{\rm bh}$ in
units of $\rm M_{\odot}$. The black hole (BH) mass function at
different redshifts is converted from the X-ray luminosity function
$\Phi(L_{\rm x},z)$ by assuming reasonable values for the
Eddington ratio $\epsilon$ and the ``duty cycle $\rm f_{\rm on}$'' of
the AGN active phase. Following the findings of various AGN
observations, $\epsilon$ depends weakly on the luminosity, where
most luminous QSOs radiate at about the Eddington limit and low
luminosity AGNs ($\rm L \,<\, 10^{44}\,\,{\rm erg/s}$) show $\epsilon \,\sim\,
0.1\,-\,0.05$. We thus approximate $\epsilon \,=\,10^{\gamma\,(log {\rm L}\,-\, 49)}$; 
$\gamma\,\approx \, 0.2$ is a scaling factor \citep{Pad89, Sal99, Wan99}. The ``duty cycle
$\rm f_{\rm on}$'', i.e. the fraction of black holes which are active
at a given time of redshift $\rm z$, may be a function of several
parameters, such as the redshift and the quasar light curves. In this work, 
we simply assume that each AGN shines for a
constant time $\rm t_{\rm Q}$, $\rm f_{\rm on}\,=\,t_{\rm Q}/t_{\rm
Hub}(z)$, where $\rm t_{\rm Hub}(z)$ is the redshift dependent Hubble
time. This means that we simplify the QSO activity as a single
significant burst with shining ratio $\epsilon$ and duration
$\rm t_{\rm Q}$ \citep{Hai00}. The QSO life time $\rm t_{\rm Q}\,=\,5\,\times\,
10^8\,\,{\rm yrs}$ is adopted in this calculation, which is close to the
e-folding time $\rm t_{\rm e}\,=\,M_{\rm BH}/\dot{M}\,=\,4\,\times\, 10^8
\,\eta/\epsilon$ shown in theoretical models with the radiation
efficiency $\eta\,\sim\, 0.1$, and consistent with the new results of
Chandra for the QSO accretion duration \citep{Wan98, Bur01,
Bar01}.

Although there are a variety of AGN spectra in the current sample
of ROSAT X-ray regime by Miyaji et al. (2000), non type 1 AGNs are
only a small fraction of the total sample and excluding them does
not change the main results significantly. In this case, we
consider all AGNs in the sample of ROSAT soft X-ray regime by
Miyaji et al. (2000) as unobscured type 1, and include a
bolometric correction $\beta$ to convert X-ray luminosity $\rm L_{\rm x}$ to black hole mass
$\rm M_{\rm bh}$. $\beta\,=\,20$ is adopted in our
calculation based on the mean type 1 AGN spectral energy
distribution from Elvis et al. (1994). The black hole mass
function $\frac{{\rm d}\,\Phi(z, M_{\rm bh})}{{\rm d}\,M_{\rm
bh}}$ can be derived from the observed X-ray luminosity function $\frac{{\rm
d}\;\Phi(z, L_{\rm x})}{{\rm d}\;L_{\rm x}}$ by:

\begin{equation}
\frac{{\rm d}\;\Phi(z, M_{\rm bh})}{{\rm d}\;M_{\rm
bh}}=\frac{0.013\,\epsilon}{\beta\,f_{\rm on}}\;\frac{{\rm
d}\;\Phi(z, L_{\rm x})}{{\rm d}\;L_{\rm x}}
\label{eq.1}
\end{equation}

A Luminosity Dependent Density Evolution(LDDE) model is suggested
by Miyaji et al.(2000), where the evolution rate drops with 
decreasing AGN luminosity, and provides a good representation of
the available X-ray observational constraints. We follow the LDDE 
cosmological evolution for the ROSAT
soft X-ray luminosity function $\frac{{\rm
d}\;\Phi(z, L_{\rm x})}{{\rm d}\;L_{\rm x}}$, and trace
the black hole mass density accreted during the active phase
with lookback time.

However, there are several lines of arguments which suggest that the optical and soft X-ray
surveys may miss a large number of type 2 AGNs, especially the high luminosity type 2 QSOs 
at high redshift ($\rm L_{\rm x}\, >\, 10^{44}\,\,erg\,\,s^{-1}$).
Synthesis models of the X-ray background in particular require a large number of luminous obscured objects to
reproduce the $2-10\,{\rm keV}$ source counts at 
relatively bright fluxes ($\,\sim\, 10^{-13}\,\,{\rm erg\,\,cm^{-2}\,\,s^{-1}}$). The ratio of type 2 to unobscured type 1 Seyferts 
($L_{\rm x}\,<\,10^{44}\,\,erg\,\,s^{-1}$) and the luminous type 2 to type 1 QSOs are 
usually in the range of $4\,-\,10$, in order to get a best fit of the available X-ray observations \citep{Mai95, Fab99, Gil01, Mai01}. 
Although the Chandra and XMM-Newton deep surveys have
recently detected several examples of the type 2 QSOs, Alexander et al.(2001) suggests that their
abundance is definitely much lower than what is expected by the XRB
population synthesis models, and probably in a ratio of about $ 1 \,\sim\, 2 $ \citep{Nor02,
Ste02, Has01, Aki02}.

Since the Miyaji et al. (2000) sample may represent only the population and evolution of the type 1 AGNs 
(i.e. type 1 Seyferts and type 1 QSOs), we adopt a reasonable fraction of obscured type 2 sources 
in our calculation to include the amount of star formation in the host galaxies of the obscured 
objects. Maiolino \& Rieke (1995) derived an estimate for the ratio of the local absorbed to 
unabsorbed low luminosity AGNs around a value of 4. Unlike the absorption distribution of local Seyferts, the
existence and abundance of the luminous type 2 QSOs is very uncertain. The difficulties are that:
1)some type 2 QSOs could be hidden in the Ultraluminous Infrared Galaxies (ULIGs) \citep{Kim98}; 2)although it is still
to be verified, some type 2 QSOs appear as normal blue, broad-lined QSOs in the optical \citep{Hal99}.

However, the present observational data suggests that the evolution of the obscured objects might be different from that of the
unobscured ones, with the number ratio of type 2 to type 1 QSOs increasing with redshift \citep{Gil99, Pom00, Ree00, Gil01}.
In this case, we divide the soft X-ray Luminosity Function(XLF) into two luminosity regions according to the $0.5\,-\,2\,\,{\rm keV}$ 
e-folding luminosity $\rm L_{\rm s}\,=\,10^{44.3}\,\, erg\,\,s^{-1}$. The abundance ratio of type 2 to type 1 
Seyferts ($\rm L_{\rm x}\,<\, L_{\rm s}$) is set
equal to 4 and the ratio of type 2 to type 1 QSOs is simplified as $\alpha\,(1+z)^p$. $\alpha$ and $\rm p$ are free parameters. 
We also explored the possibility that the abundance ratio of LLAGNs is much higher than 4, i.e. as high as 10. It seems that 
the current far-infrared and submillimeter deep surveys are not robust
enough to contrain this value. Probably, it is because LLAGNs do not contribute much to the bright SCUBA number counts. In this case, the abundance ratio of the low luminosity Seyferts is not critical for this work.
Thus, the abundance ratio of type 2 to type 1 AGNs including low luminosity Seyferts and QSOs can be described as the following:

\begin{equation}
R_{2-1}=4 \, e^{-\frac{L_{\rm x}}{L_{\rm s}}}+\alpha\,(1+z)^p\,(1 -e^{-\frac{L_{\rm x}}{L_{\rm s}}})
\end{equation}

\begin{figure}
\plotone{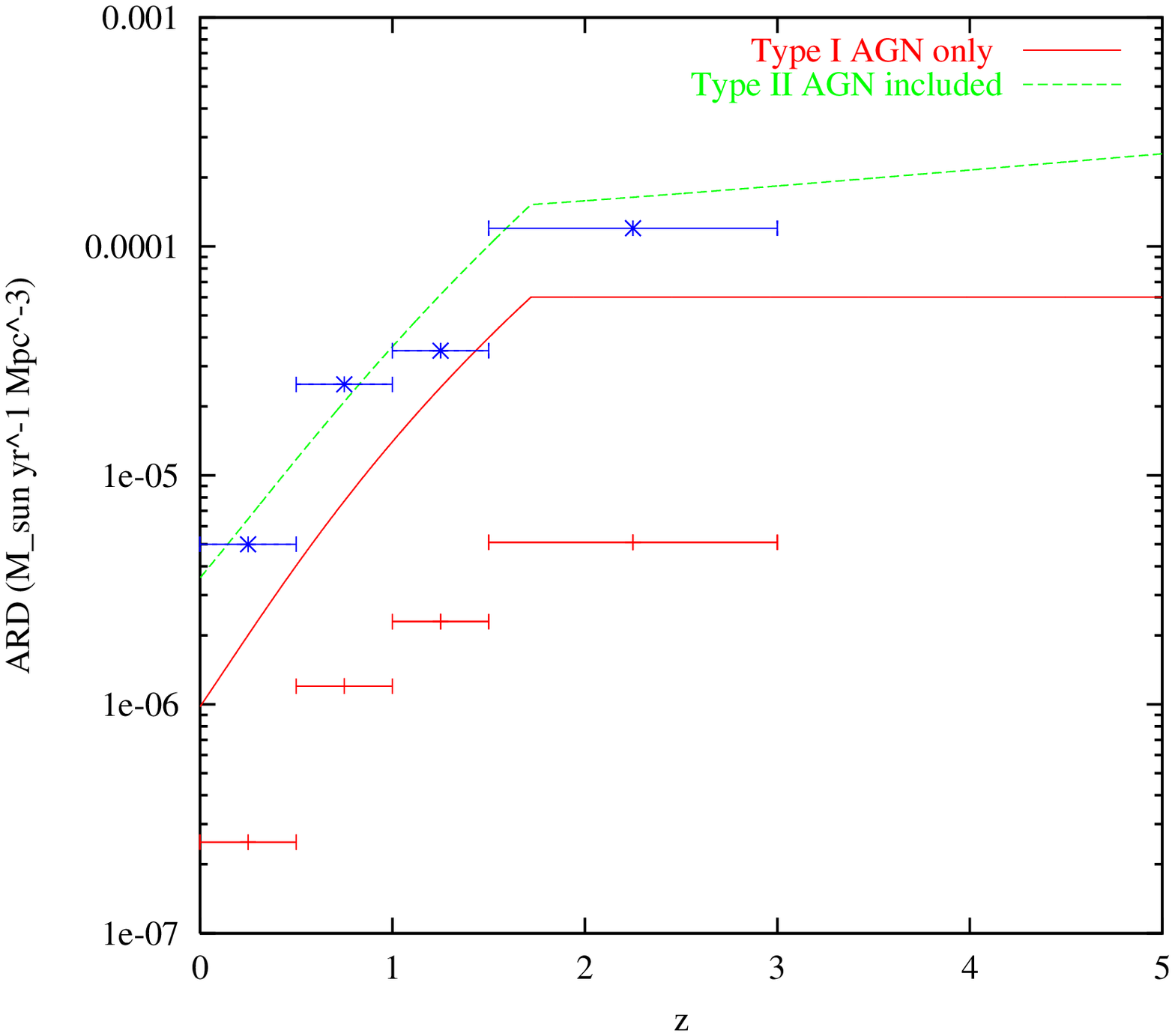}\caption{Accretion rate density ($\rm ARD$) in units of $\rm  M_{\odot}\,\,yr^{-1}\,\,Mpc^{-3}$ vs. redshift from the model 
calculation. The solid line illustrates the calculated accretion history of type 1 AGNs, while the dashed line is for the case
with a reasonable fraction of type 2 AGNs included. The data points, blue asterisks 
represent the accretion density calculated from the bolometric luminosities of 69 hard X-ray
selected sources; while the red crosses are calculated from their X-ray luminosities as a 
low limit \citep{Bar01}.\label{fig1}}
\end{figure}

Considering the present day black hole density
$\rho_{\rm BH}\,\sim\, 3\,-\,5\,\times\, 10^5\,\,{\rm M_{\odot}/Mpc^3}$ and the submillimeter deep
surveys as the model constraints, we found the ratio of the obscured type 2 to
unobscured type 1 QSOs has an upper limit of about 2 ($\alpha$ and $\rm p$ have the best fit values of $0.1$ and $1.9$). 
There is not much room left for a ratio far beyond this number.
This is actually consistent with
the results of recent Chandra deep surveys by Alexander et al.
(2001) who suggests that this ratio is unlikely to be more than 8 and
is probably considerably lower. Since X-ray emission directly measures the accretion of AGNs, 
we show in Fig. \ref{fig1} the accretion history from the model calculation. The solid
line illustrates the accretion 
history of the type 1 AGNs only; while the dashed line is for the case
with type 2 AGNs included. The
accretion rate density $\dot{\rho_{\rm bh}}$ in units of $\rm M_{\odot}\,yr^{-1}\,Mpc^{-3}$ from
a multiwavelength study of 69 hard X-ray selected sources in the field of Abell 370 cluster (A370), Hawaii Survey Field (SSA13)
and Chandra Deep Field-North (CDF-N) for four redshift bins are also ploted in Fig. \ref{fig1}, where the 
blue asterisks represent the accretion
onto the AGNs calculated from their bolometric luminosities and the red crosses are from their X-ray 
luminosities as a low limit. Fig. \ref{fig1} shows that the predicted accretion rate 
density of the type 1 AGNs is below the upper bound
derived from the hard X-ray selected samples \citep{Bar01}. Meanwhile, we see that the fraction of type 2 AGNs included 
in our model still agrees with the current results of deep hard X-ray surveys. It is worth mentioning that the
treatment of the abundance of type 2 AGNs in this work is very rough. We need more data from the Chandra and XMM-Newton 
surveys at significant fainter fluxes to better constrain the abundances.

\section{Star formation associated with black hole growth and the SCUBA counts}

Recent observations seem to support an evolutionary
scenario where the black hole growth and intensive star
formation during the spheroid formation are co-evolving events.
In this scheme, the accretion onto a seed black hole and the rapid
star formation in a $\,\sim\, {\rm kpc}$ region are significantly enhanced due
to a merging process or tidal interactions. Accretion $+$ star formation
compete for the gas supply and may self-regulate by the
fundamental properties of the viscous accretion disk, leading to
the formation of massive black holes scaling with their spheroids. 
In this case, the mass function of spheroids may have a
similar form to the balck hole mass function, and the spheroidal 
mass distribution $\frac{{\rm d}\;\Phi(z, M_{\rm
sph})}{{\rm d}\;{\rm M_{\rm sph}}}$ could be derived by:

\begin{equation}
\frac{{\rm d}\;\Phi(z, M_{\rm sph})}{{\rm d}\;M_{\rm sph}}=\frac{{\rm d}\;\Phi(z, M_{\rm bh})}{{\rm d}\; M_{\rm bh}}\,\centerdot\,{\rm 
R(M_{\rm bh}/M_{\rm sph}})
\label{eq.2}
\end{equation}

\noindent $\rm R(M_{\rm bh}/M_{\rm sph})$ is the black hole to bulge
mass ratio. We adopt here a mean value $\rm R(M_{\rm bh}/M_{\rm
sph})\,\sim\, 0.002$ as the first approximation \citep{Mcl94, Kor95,
Fab97, Mag98, Mer01a}.

The aim of this section is mainly to estimate the star formation rate
during spheroid formation in individual galaxies, and then discuss
the co-moving star formation history related with such an epoch.
The SCUBA number counts are used as a model constraint or
consistency check for the amount of star formation, especially at
$z \,>\, 1$.

The basic assumption of this calculation is that the stellar
population of the spheroids are mostly from an intensive
star formation phase in the central $\rm kpc$ region which is associated
with the mass accretion onto the central black holes. In the model
calculation of Wang \& Biermann (1998), star formation would
dominate the energy output in the first $\,\sim\, 5\,\times\, 10^8\,\,{\rm yrs}$
for the stellar black hole growing to a mass of $\,\sim\,
10^6\,\,{\rm M_{\odot}}$, even with full-Eddington accretion. Because of
the heavy dust extinction, the intensive star formation during
this process would emit a large fraction of its energy at
infrared wavelength. The Ultraluminous Infrared Galaxies (ULIGs) may represent such a
dust-enshrouded stage of both intensive starbursts and AGNs. Once
the AGN activity becomes powerful enough, the nucleus sheds its
obscuring dust, allowing the AGN to visually dominate, completing the
evolutionary process from ULIGs to optically selected
QSOs \citep{San88, Yam94, Mih96, Sil98, Tan99}.

We adopt a characteristic star formation time scale $\tau_{\rm
sf}$ for the starburst
activities coupled with the black hole growth, which is approximately equal to the
QSO life time $\rm t_{\rm Q}$ (i.e. $\tau_{\rm
sf}\,\sim\, t_{\rm Q}\,\sim\, 5\,\times\, 10^8\,\,{\rm yrs}$). This is consistent with
the predictions of the theoretical models and the recent Chandra results of the
QSO accretion duration \citep{Wan98, Bur01, Bar01}. The star
formation rate during such a phase would be $\rm SFR \,=\,M_{\rm
sph}/\tau_{\rm sf}$. Following the cosmological evolution inferred
from X-ray deep surveys, we get the redshift dependent black hole
mass distribution by eq. \ref{eq.1}, and the mass function of spheroids
with redshift by eq. \ref{eq.2}. So far, we get to know the star
formation rate in individual spheroid during the past active
phase, the mass distribution and the cosmological evolution. The
co-moving star formation rate vs. redshift from our calculation is
shown in Fig. \ref{fig2}.

In section \ref{sec2}, we have applied the local black hole
density to constrain the black hole mass distribution
and cosmological evolution in our calculation.
Independently, submillimeter deep survey would be a good tracer
and constraint of such an active star formation stage associated
with the major phase of black hole growth and spheroid formation.
To discuss the contribution of such an intensive star formation
phase to the SCUBA number counts, we should convert the star
formation rate to far-infrared emission for individual source.
This relation is not straightforward even for the local ULIGs because
of the high internal extinction. A number of authors have
discussed how the star formation rate in a galaxy can be inferred
from its optical, UV or far-infrared luminosity \citep{Sco83,
Lei95, Row97}. The transformation factor depending upon star
formation parameters could vary by more than one order of magnitude.
We adopt a mean value $\rm L_{\rm FIR}/L_{\odot}\,=\,3.8 \,\times\, 10^9\,\,SFR\,\,
({\rm M_{\odot}\,yr^{-1}})$, and a mean color ratio $\rm R_{\rm c}\,=\,L_{\rm
FIR}/L_{850}\,\sim\, 5\,\times\, 10^3$ given by Chary \& Elbaz (2001) from IRAS,
ISO and SCUBA surveys.

Similar to eq. \ref{eq.1},  we can derive the $850\,\mu
m$ luminosity function at different redshift as follow:

\begin{equation}
\frac{{\rm d}\;\Phi(z, L_{850})}{{\rm d}\;L_{850}}= f_{\rm
\widehat{on}}\;\frac{{\rm d}\;\Phi(z, M_{\rm sph})}{{\rm d} \;M_{\rm
sph}}\;\frac{{\rm d}\;M_{\rm sph}}{{\rm d} \;L_{850}}
\label{eq.3}
\end{equation}

\noindent where $\rm f_{\rm \widehat{on}}\,=\,t_{\rm sf}/t_{\rm Hub}$ reflects the
fraction of galaxies which are in the active stage with intensive
star formation ongoing. $\frac{{\rm d}\,{\rm M_{sph}}}{{\rm
d}\,{\rm L_{850}}}\,=\,\frac{\rm \nu\,\,R_{\rm c}\,\, {\rm t_{\rm sf}}}{3.8
\,\times\, 10^9\,L_{\odot}}$, $\rm R_{\rm c}$ and $\rm t_{\rm sf}$ are the parameters
adopted in our model and discussed above. $\nu \,=\,3.5\,\times\, 10^{11}$
is the frequency of $850\,\mu m$. Since early star formation could be well
constrained by the far-infrared and submillimeter deep surveys, we show the $850\,\mu m$ number count
fitting in Fig. \ref{fig3}. We find
that the amount of star formation related to the spheroid formation during the black
hole growth heats the dust sufficiently to account for the
far-infrared emission in most of the SCUBA counts.

Throughout the discussion above, we only focus on the star
formation history in massive spheroids, i.e. elliptical galaxies
or the disk galaxies with bulges. Although there are two kinds of bulges in the
disk galaxies strengthened by especially Hubble Space Telescope ("mini-elliptical"  vs. "pseudobulge") 
and their formation mechanism might be very different, both of them follow the
same correlation of $\rm M_{\rm bh}\,-\,M_{\rm B, bulge}$ or $\rm M_{\rm
bh}\,-\,\sigma_{\rm e}$, presumably consistent with a scenario
that the bulge formation and the black hole growth are closely
connected \citep{Car97, Car98a, Car98b, Pel00}. The star
formation during the massive spheroid formation would be drastic
and dusty, very probably missing in the current UV/optical deep surveys.
However, some disk galaxies harbor only small black holes with the black hole to bulge
mass ratio much lower than that of the spheroidal systems. The typical
example is M33, a bulgeless galaxy with the upper limit on a black
hole mass $\rm M_{\rm bh}\,\le\, 1000\,M_{\odot}$ (three magnitude lower
than what is expected from the mass correlation in massive systems). The
star formation in these galaxies may be mild and the dust
extinction is not so severe. Normally they are contained in the
optical deep surveys, or other studies of the spiral
galaxies \citep{Cav00, Hor02}. The detailed discussion of the
star formation history of these galaxies are beyond
the scope of this paper.

\section{Discussion and summary}

We have begun to probe the cosmic star formation
history associated with AGN accretion by X-ray deep surveys based
on the tight correlation of $\rm M_{\rm bh}\,-\,M_{\rm B, bulge}$ in early
type galaxies and nearby AGNs. This approach is parallel and
complementary to the current study from optical and infrared
observations of the star formation activities at $\rm z\,>\,1$ \citep{Mad98, Pet98, Hug98, Bar98, Ell98}.

\begin{figure}
\plotone{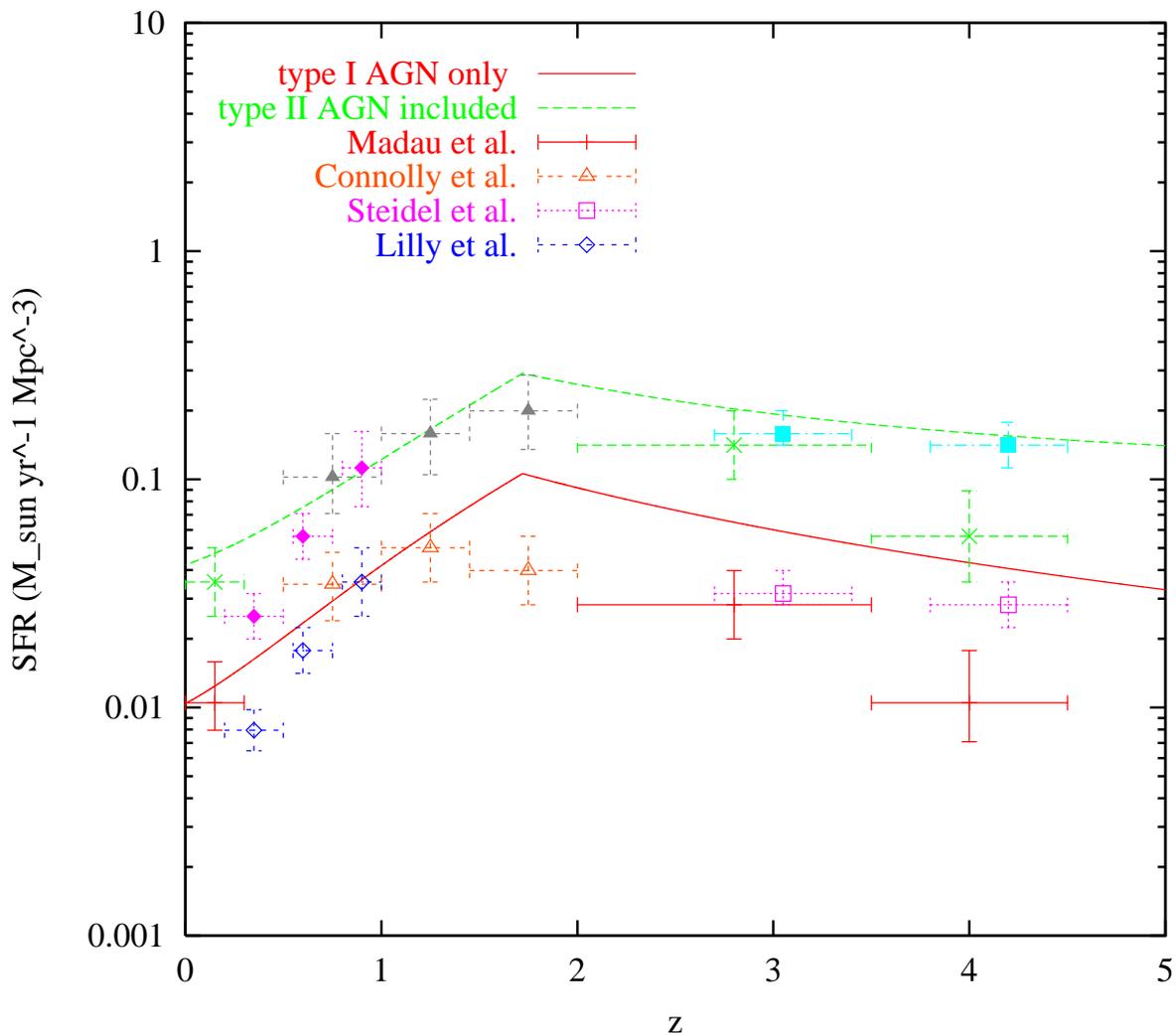} \caption{Co-moving star formation rate vs.
  redshift predicted by the model.  The data sources are
  indicated in the Figure, while
  the filled symbols are those with dust correction ($\times 4.7$) adopted by
  Steidel et al. 1999. The upper line (green) is the star formation in the host galaxies
  of whole AGN sample including type 1 and a reasonable fraction of type 2 AGNs, while the lower line (red) is the contribution from type 1 AGNs only. \label{fig2}}
\end{figure}

The cosmic star formation history associated with AGN accretion
derived from our calculation is approximately comparable to that
of the normal galaxies in Madau plot after a reasonable dust correction,
although they are much rarer objects compared with optically
selected galaxies (The co-moving number density of luminous AGNs is as small as 
$10^{-7}\,\,{\rm Mpc^{-3}}$, see Fig. \ref{fig2}). In this case, we might say that about half
of the star formation (if not more) in our Universe is closely
connected with AGN accretion, and we might severely overlook
the intensive star formation during the epoch of
the spheroidal formation by optical/NIR surveys due to the significant dust extinction and the small
sample volume of these surveys. 

We found from Fig. \ref{fig2} that the peak of the intensive star formation
representing the spheroid formation is at $\rm z\,\sim\, 2$, not
necessarily much beyond this epoch even if we take into account a
reasonable fraction of type 2 QSOs at high redshift. According to various observation, we divide the soft
X-ray luminosity function into two luminosity regions, where the abundance ratio of
the type 2 to type 1 Seyferts is set equal to 4, and the ratio of the type 2 to type 1
QSOs is simplified as a power law function of $\alpha\,(1+z)^{p}$. Actually an upper limit of the abundance ratio
of the type 2 to type 1 QSOs is about $1\,\sim \, 2$ from our
calculation with the constraints of the local black hole mass
density and the results of the submillimeter deep surveys. However, current model
constraints are not sufficient to reject a much higher abundance ratio for low luminosity
type 2 to type 1 AGNs ($>4$). The reason may be that the bright SCUBA counts are dominated
by the luminous AGNs and the ratio of the low luminosity AGNs is not critical. Future results of the Chandra and
XMM/Newton deep surveys would give more information on the abundance of the obscured objects.

\begin{figure}
\plotone{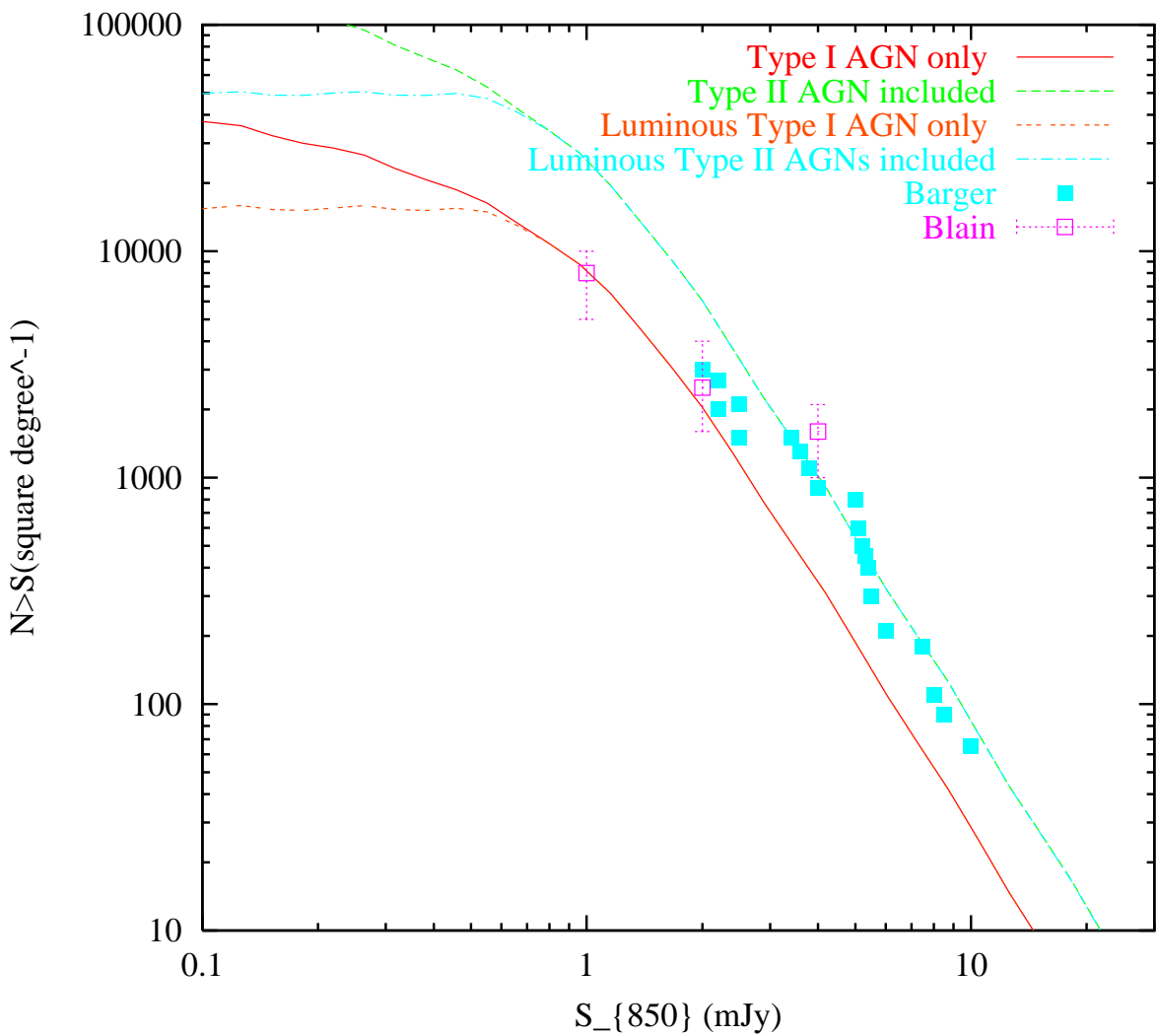} \caption{$850\,\mu m$ number count fitting from
the model calculation in case of type 1 AGNs only and the case of type 2 AGNs included. The data are from Barger et al. (1999) and Blain et al. (1999). We see from the figure that
the star formation powered far-infrared emission from X-ray luminous AGNs ($L>10^{44}\,erg/s$) would dominate the SCUBA counts,
while the faint parts ($\rm S_{850} < 1\,mJy$) might be from those low luminosity X-ray sources.\label{fig3}}
\end{figure}

The energy budget of the submillimeter sources has been discussed by several authors, 
which show that the AGN powered far-infrared emission in the obscured objects could account
for only a certain fraction of the SCUBA number counts \citep{Alm99,
Ris02}. The SCUBA/X-ray anti-correlation
of the Chandra deep surveys gives a clear detection
of AGN activities in those samples only $\,\sim\, 10\%$ \citep{Bau00, Fab00, Bar01b, Hor01}. Nevertheless, Fig. \ref{fig3} shows that the
far-infrared emission from the dust heated by the intensive star formation
during the black hole growth could sufficiently interpret the number counts of the
submillimeter deep surveys, where the SCUBA number counts are dominated by the contribution
from the star formation in the host galaxies of the luminous type 1 or type 2 QSOs, and 
the the faint parts ($\rm S_{850}< 1\,mJy$) may be from those low luminosity X-ray sources with small bulges. In this case, our calculation may indicate that the star formation activity might
dominate the energy power (at least comparable to the AGNs) in the far-infrared
emission in these SCUBA sources, consistent with the multiwavelength observations of the 
submillimeter selected galaxies. They suggest that even when an AGN is present in a SCUBA
source, it rarely dominates the engergy budget of the galaxy \citep{Fra98, Ale02, Ivi02, Sma02}.

\acknowledgments
  YPW acknowledges the COE Fellowship of Japan and the NSFC 10173025 of
  China. We would like to thank the anonymous referee for the helpful
  comments and suggestions. YPW also feels very grateful to Ms Jenny 
  Greene for her kindness to correct the English of this paper.

\clearpage


\end{document}